\documentstyle[11pt,IAUS212,twoside,epsf]{article}

\def\hii{H{\sc ii}}
\def\mabs{$M_{\rm B}$}

\def\oii{[O\,{\sc ii}] $\lambda\lambda$3726,3728}
\def\oiii{[O\,{\sc iii}] $\lambda\lambda$4959,5007}
\def\nii{[N\,{\sc ii}] $\lambda$6584}
\def\halpha{\ifmmode {\rm H{\alpha}} \else $\rm H{\alpha}$\fi}
\def\hbeta{\ifmmode {\rm H{\beta}} \else $\rm H{\beta}$\fi}
\def\zsun{Z$_{\odot}$}

\def\edcomment#1{\iffalse\marginpar{\raggedright\sl#1\/}\else\relax\fi}
\reversemarginpar

\begin{document}
\title{Physical Properties of Low-Luminosity Galaxies at $z \sim 2$}
 \author{M. Lemoine-Busserolle, T. Contini, R. Pell\'o, J.-F. Le Borgne, \& J.-P. Kneib}
 \affil{LAOMP - UMR 5572, 14 avenue E. Belin, F-31400 Toulouse, France}
 \author{C. Lidman}
 \affil{European Southern Observatory, Alonso de Cordova, 3107 Vitacura, Chile}

\begin{abstract}
We report the results obtained from ISAAC/VLT near-IR spectroscopy of 
two low-luminosity $z \sim 1.9$ galaxies located in the core 
of the lensing cluster AC114. The amplification factor allowed to 
obtain, for the first 
time, physical properties (SFR, abundance ratios, mass, etc) of
 star-forming galaxies, 1 to 2 magnitudes fainter than in previous 
studies of LBGs at $z \sim 3$.
\end{abstract}

Near-infrared (NIR) spectroscopy provides a view of the physical 
properties of the high-redshift (up to $z \sim 6$) star-forming galaxies 
using various rest-frame optical line diagnostics (Balmer lines, \oii\, 
\oiii\, \nii, etc) commonly applied for nearby galaxy studies. With the recent 
advent of NIR spectrographs on 8-10m class telescopes, these studies have 
started in the $1.8 \leq z \leq 3.5$  domain (Pettini et al. 1998, 2001; 
Kobulnicky \& Koo 2000). The main applications have been the determination of 
the star formation rate (SFR) deduced from the \halpha\ luminosity, the 
abundance of oxygen derived from the strong oxygen emission lines, 
and virial masses obtained from the emission line widths. 
The main goal of these studies is to derive the physical properties of 
distant galaxies using the same parameter space as for galaxies in the local 
Universe. 

We report, in this contribution, the results obtained on two lensed 
low-luminosity $z \sim 1.9$ galaxies (S2 and A2) located in the core of 
the lensing cluster AC114. The amplification factor allowed to obtain, 
for the first time, emission-line measurements on star-forming galaxies 
as faint as \mabs\ $\sim -20$, thus 1 to 2 magnitudes fainter than in 
previous studies of LBGs at $z \sim 3$ (e.g. Pettini et al. 2001). 
ISAAC/VLT medium-resolution $JHK$ spectra of the two galaxies S2 and 
A2 have been obtained in two hour exposure (per band and per object) 
on 26-27 september 2001 and reduced in a standard way. Bright rest-frame 
optical emission lines are measured (intensity and line width) with a 
high S/N ratio (see Figure 1) and have been used to perform a detailled 
analysis of the physical properties (SFR, chemical abundances, 
virial mass, etc) of S2 and A2. 
The SFRs derived from \halpha\ luminosity are 20 (S2) and 40 (A2) times 
higher than those derived from the UV (1500\AA) luminosity, without 
dust extinction correction. The difference reduces to a factor of 3 (S2) 
and 6 (A2) assuming an extinction E(B-V)=0.3 mags, but still the SFR(\halpha) 
$>$ SFR(1500\AA). These results suggest that quite large dust extinction 
corrections are needed for intrinsically faint optically-selected galaxies 
compared to other samples of bright UV-selected distant galaxies (e.g. 
Pettini et al. 2001).
The large spectral coverage obtained, from \oii\ to \halpha,\nii\, allowed 
to set strong constraints on the abundances ratios of these two low-luminosity 
galaxies (see Figure 1). The behavior of S2 and A2 in terms of metallicity 
is very different, and they are also different from typical LBGs at 
$z \sim 3$. S2 is a low-metallicity object (Z $\sim 0.03$ \zsun) with a 
low N/O ratio, similar to those derived in the most metal-poor nearby 
\hii\ galaxies. On the contrary, A2 is a metal-rich galaxy (Z $\sim 1.5$ 
\zsun) with a high N/O abundance ratio, similar to those derived in the most 
metal-rich massive SBNGs. The position of MS 1512-cB58, a lensed luminous 
LBG (Teplitz et al. 2000), is intermediate between these two extremes 
showing abundance ratios typical of low-mass SBNGs and intermediate-redshift 
galaxies. These results suggest different star formation history for 
distant galaxies of different luminosities. Larger samples of faint 
high-$z$ galaxies are urgently needed to conclude, in order to set strong 
constraints on the galaxy evolution scenario.\\ 

\begin{figure}[t]
\plottwo{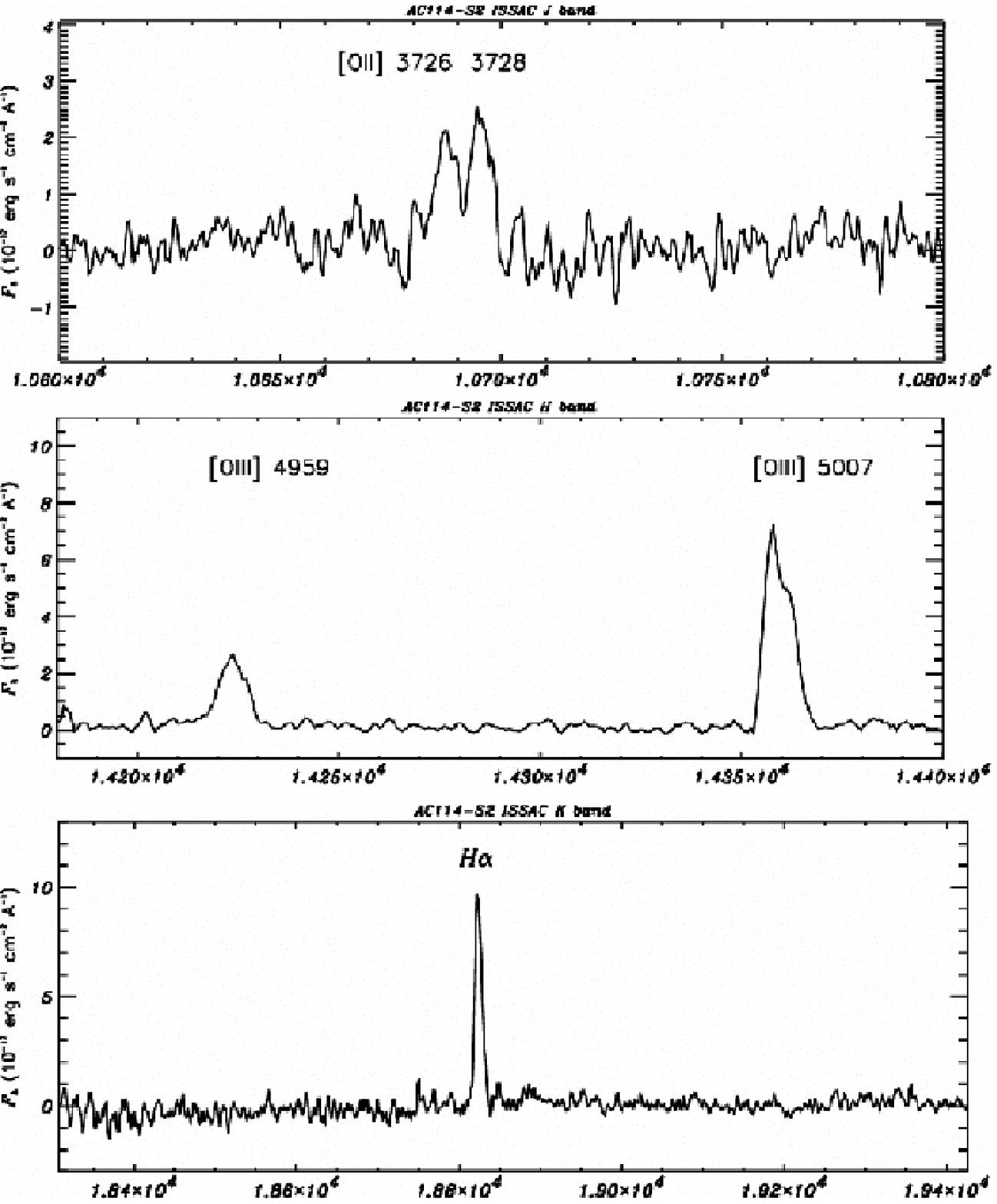}{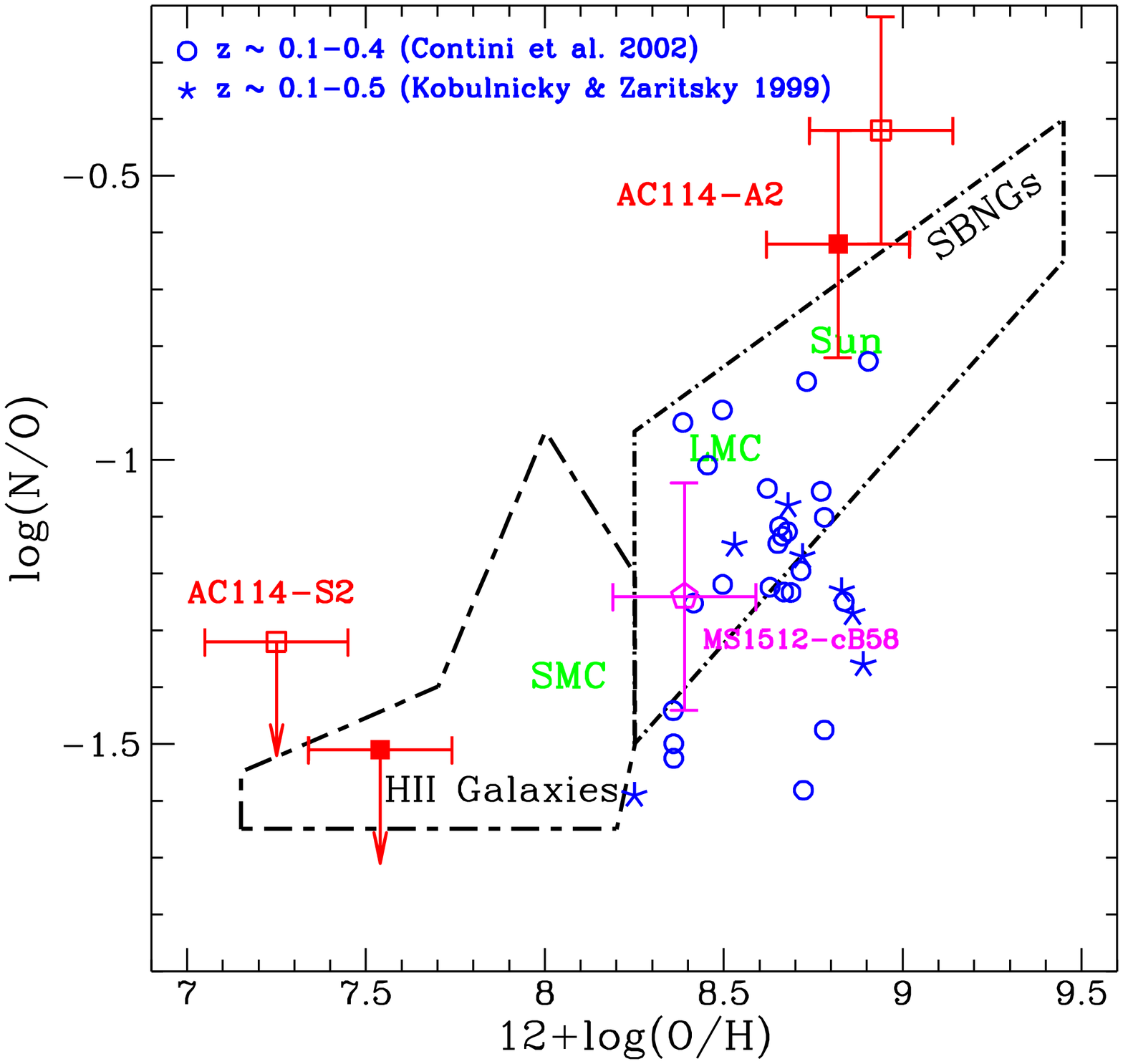}
\caption{{\it Left:} ISAAC/VLT spectrum of the low-luminosity 
$\sim 1.9$ galaxy S2 in the core of the lensing cluster AC114. 
{\it Right:} N/O versus O/H relation for nearby (\hii\ galaxies 
and SBNGs) and high-$z$ star-forming galaxies. The location of 
the two low-luminosity $z \sim 1.9$ galaxies S2 and A2 is shown 
without extinction correction (empty squares) and assuming E(B-V)=0.3 
(filled squares). 
}
\end{figure}
\vspace{-0.3cm}
\begin{acknowledgements}
Part of this work was supported by a {\it ``Conseil R\'egional de la Martinique''} grant and 
the french CNRS. 
\end{acknowledgements}


\vspace{-0.6cm}
\begin{references}
\vspace{-0.3cm}

\reference Contini, T., Treyer, M.-A., Sullivan, M., Ellis, R.S. 2002, MNRAS 330, 75


\reference Kobulnicky, H.~A., Koo, D.~C. 2000, ApJ 545, 712

\reference Kobulnicky, H.~A., Zaritsky, D. 1999, ApJ 511, 118


\reference Pettini, M., Kellogg, M., Steidel, C.~C., et al. 1998, ApJ 508, 539

\reference Pettini, M., Shapley, A.~E., Steidel, C.~C., et al. 2001, ApJ 554, 981

\reference Teplitz, H.~I., McLean, I.S., Becklin, E.E., et al. 2000, ApJ, 533 L65

\end{references}
\end{document}